\pdfoutput=1
\documentclass[%
reprint,
superscriptaddress,
 amsmath,amssymb, 
 aps,
 prl,
floatfix,
]{revtex4-2}

\usepackage{graphicx}
\usepackage{epsfig} 
\usepackage{dcolumn}
\usepackage{bm}
\usepackage{epstopdf}
\usepackage{multirow}
\usepackage{amsmath}
\usepackage{booktabs}
\usepackage{color}
\usepackage{gensymb}
\usepackage{kantlipsum}  
\usepackage{hyperref}
\usepackage{ulem}
\usepackage{braket}
\usepackage{physics}
\usepackage[utf8]{inputenc}
\usepackage[T1]{fontenc}
\usepackage{mathptmx}

\begin{document}

\title{Collisionally Stable Gas of Bosonic Dipolar Ground State Molecules}

\preprint{APS/123-QED}

\author{Niccol\`{o} Bigagli}
\affiliation{Department of Physics, Columbia University, New York, New York 10027, USA}
\author{Claire Warner}
\affiliation{Department of Physics, Columbia University, New York, New York 10027, USA}
\author{Weijun Yuan}
\affiliation{Department of Physics, Columbia University, New York, New York 10027, USA}
\author{Siwei Zhang}
\affiliation{Department of Physics, Columbia University, New York, New York 10027, USA}
\author{Ian Stevenson}
\affiliation{Department of Physics, Columbia University, New York, New York 10027, USA}
\author{Tijs Karman}
\affiliation{Institute for Molecules and Materials, Radboud University, 6525 AJ Nijmegen, Netherlands}
\author{Sebastian Will}\email{Corresponding author. Email: sebastian.will@columbia.edu}
\affiliation{Department of Physics, Columbia University, New York, New York 10027, USA}

\date{\today}

\begin{abstract}

Stable ultracold ensembles of dipolar molecules hold great promise for studies of many-body quantum physics, but high inelastic loss rates have been a long-standing challenge. Recently, gases of fermionic molecules in their ground state have been effectively stabilized by applying external fields. However, for gases of bosonic molecules, which might provide access to fundamentally different many-body quantum systems, it is unknown whether a similar suppression of losses can be achieved. This is due to the high inelastic loss rates for bosonic molecules, which are intrinsically one to two orders of magnitude larger than for their fermionic counterparts. Here, we stabilize a bosonic gas of strongly dipolar NaCs molecules via microwave shielding, decreasing losses by more than a factor of 200 and reaching lifetimes on the order of 1 second. In addition, we measure high elastic scattering rates and characterize their anisotropy, which arises from strong dipolar interactions. Finally, we demonstrate evaporative cooling of a bosonic molecular gas. We increase the phase-space density by a factor of 20, reach a temperature of 36(5) nK, and bring the system to the brink of quantum degeneracy. Our results constitute a step towards the creation of a Bose-Einstein condensate of dipolar molecules and open the door to the creation of strongly correlated phases of dipolar quantum matter.

\end{abstract}

 \maketitle

\section{Introduction}

Ultracold gases of atoms and molecules have revolutionized the experimental exploration of many-body quantum systems \cite{bloch2008many,carr2009cold,gross2017quantum,bohn2017cold}. In recent years, systems with dipolar long-range interactions have gained rapid traction, enabling the creation of new types of strongly correlated and highly entangled quantum matter. Magnetic atoms  \cite{lahaye2009physics,chomaz2022dipolar}, operating in a regime of relatively weak dipole-dipole interactions, have allowed the realization of quantum ferrofluids \cite{ kadau2016observing} and the creation of droplet \cite{schmitt2016self} and supersolid phases \cite{norcia2021two}. Rydberg atoms \cite{browaeys2020many}, operating in a regime of extremely strong interactions, have given rise to phases with crystalline order \cite{schauss2015crystallization}, simulation of quantum magnetic models \cite{labuhn2016tunable, semeghini2021probing}, and the controlled creation of entanglement \cite{levine2019parallel}. Dipolar molecules \cite{baranov2012condensed} are expected to operate in an intermediate regime where kinetic energy and interaction energy are on a similar scale, giving rise to nontrivial correlations and self-organization \cite{lahaye2009physics}. Predicted phases include strongly interacting superfluids \cite{wang2006quantum,cooper2009stable}, supersolids \cite{trefzger2009pair,schmidt2022self}, dipolar crystals \cite{buchler2007strongly,rabl2007molecular}, and Mott insulators with fractional filling \cite{Capogrosso2010}. However, molecules have been found to be prone to strong inelastic loss \cite{ospelkaus2010quantum,julienne2011universal,bause2023ultracold}. Suppression of losses is a critical prerequisite to realize stable many-body quantum systems of dipolar molecules.

Quantum statistics plays an important role in molecular loss dynamics \cite{quemener2012ultracold}. Fermionic molecules are intriniscally less prone to inelastic loss than bosonic ones. For indistinguishable fermions, the probability of reaching short range in a two-body collision is suppressed by the $p$-wave centrifugal barrier \cite{ni2008high,park2015ultracold,de2019degenerate}. For bosons, such a barrier is absent and the rate of two-body loss is typically one to two orders of magnitude larger than for fermions \cite{julienne2011universal}. To reduce loss below the natural rate, shielding techniques have been proposed that utilize external electric fields to engineer a repulsive barrier for intermolecular collisions, leveraging the rich internal state structure of molecules \cite{avdeenkov2006suppression,gorshkov2008suppression,gonzalez2017adimensional,karman2018microwave,lassabliere2018controlling}. Microwave shielding was demonstrated in a proof-of-principle experiment for two bosonic CaF molecules in an optical tweezer trap \cite{anderegg2021observation}. For bulk gases of fermionic molecules, shielding with d.c.~electric fields was shown for KRb \cite{valtolina2020dipolar,matsuda2020resonant,li2021tuning} and microwave shielding for NaK~\cite{schindewolf2022evaporation}, suppressing inelastic loss by about an order of magnitude, sufficient to demonstrate evaporative cooling. Whether loss in bosonic molecular gases can be sufficiently suppressed to enable evaporative cooling has remained an open question. 

Here, we demonstrate the stabilization of a gas of bosonic sodium-cesium (NaCs) molecules against inelastic loss via microwave shielding. NaCs is strongly dipolar with a permanent dipole moment of $d_0 = 4.75(20)$~D \cite{dagdigian1972molecular}, making shielding highly effective.  We observe the suppression of two-body loss by more than a factor of 200, increasing the lifetime of dense ensembles, with an interparticle spacing of about 1 $\mu$m, from 16(2) ms to 1.0(1) s. The microwave field induces a dipole moment of up to 1.3 D in the laboratory frame, leading to significant dipolar interactions that enhance elastic collisions. Via cross-thermalization we measure strong elastic interactions and obtain a ratio of elastic-to-inelastic collisions of up to $\gamma = 4 (1) \times 10^3$. Under these conditions, we demonstrate evaporative cooling of a bosonic molecular gas, increasing its phase-space density (PSD) by a factor of 20 and reaching a temperature of $36(5)$ nK.

\begin{figure*}
    \centering
    \includegraphics[width = \textwidth]{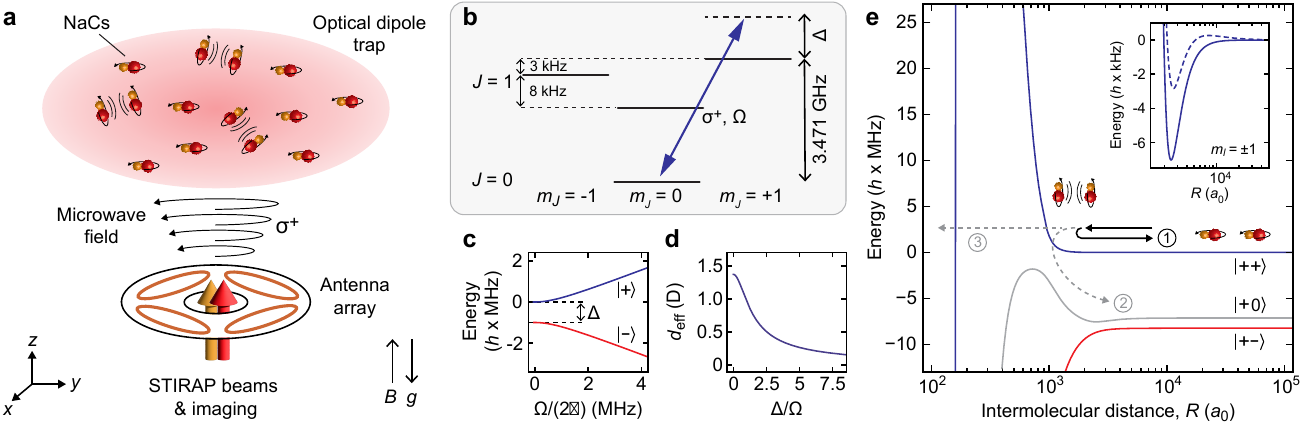}\\
    \caption{Microwave shielding of NaCs molecules. \textbf{a}, Illustration of the trapped molecular gas. Molecular dipoles are set into rotation by the electric field of a $\sigma^+$-polarized microwave field generated by an antenna array. Vertical beams for stimulated Raman adiabatic passage (STIRAP) allow for time-of-flight expansion of the microwave-shielded gas up to 50 $\mu$m cloud waist, enabling precise measurement of temperature. \textbf{b}, Rotational levels of NaCs at 864 G. The states $|J, m_J\rangle = |0, 0\rangle$ and $|1, 1\rangle$ are split by an energy $\hbar \omega_{\mathrm{res}}$ with $\omega_{\mathrm{res}}= 2 \pi \times 3.471323(2)$ GHz. The microwave field is blue-detuned with respect to the resonance by an amount $\Delta$. \textbf{c}, Energy of the single-molecule dressed states, $|+\rangle$ and $|-\rangle$, as a function of Rabi frequency. The states are split by an energy $\hbar \Omega_\mathrm{eff} = \hbar \sqrt{\Omega^2 + \Delta^2}$. \textbf{d}, Effective dipole moment in the lab frame of the $|+\rangle$ state as a function of $\Delta / \Omega$. \textbf{e}, Potential energy curves of a pair of microwave dressed molecules approaching in the $s$-wave channel for $\Omega / (2 \pi) = 4$ MHz and $\Delta / (2 \pi) = 6$ MHz. The adiabatic potentials for $|++\rangle$ (blue solid line), $|+0\rangle$ (grey solid line), and $|+-\rangle$ (red solid line) are shown. Molecules are either (1) reflected by the repulsive potential, (2) lost to non-shielded states, or (3) reach short range. The inset shows the difference between bosonic NaCs ($s$-wave scattering, solid line) and a hypothetical fermionic NaCs molecule ($p$-wave scattering with $m_l = \pm 1 $, dashed line) for which the $p$-wave barrier provides further shielding.  \\} 
    \label{fig:1}
\end{figure*}

\section{Shielding}

Collisional stabilization is achieved by exposing the ultracold molecular gas to a microwave field with specifically chosen polarization, frequency, and intensity (Fig. 1a). Initially, the molecules are in the rotational ground state $|J, \ m_J\rangle = |0, \ 0\rangle$, where $J$ denotes the total angular momentum excluding nuclear spin and $m_J$ its projection onto the quantization axis. Then, a circularly polarized microwave field is applied at a frequency that is blue-detuned by an amount $\Delta$ from the resonance $\omega_\mathrm{res}$ with the excited state $|J, \ m_J\rangle = |1, \ 1\rangle$ (Fig. 1b). The intensity of the microwave field is adiabatically increased, transferring each molecule into the state $|+\rangle = \cos(\phi) |0, \ 0\rangle + \sin(\phi) |1, \ 1\rangle$, where the mixing angle  $\phi$ is defined by $\sin(2 \phi) = 1 / \sqrt{1 + (\Delta / \Omega)^2}$ and $\Omega$ denotes the Rabi frequency (see Methods). The orthogonal dressed state, $|-\rangle = \sin(\phi) |0, \, 0\rangle - \cos(\phi) |1, \, 1\rangle$, remains unpopulated. Fig. 1c shows the energy splitting between the dressed states as a function of Rabi coupling.  

In a semiclassical picture, the dressed states represent dipoles rotating in the $xy$-plane at a frequency $\omega_{\mathrm{res}} + \Delta$. Due to the superposition of opposite parity states $|0,0\rangle$ and $|1,1\rangle$, the dressed states feature an induced dipole moment. The effective dipole moment, $d_\mathrm{eff}$, as a function of $\Delta/\Omega$ is shown in Fig. 1d.  At large intermolecular distances, shielded molecules interact through the long-range dipole-dipole interaction $V_{\rm dd} = d_\mathrm{eff}^2 (3 \cos^2\theta-1)/(4 \pi \epsilon_0 R^3)$ \cite{yan2020resonant}, where $\theta$ denotes the angle between the rotation axis of the dipole and the intermolecular axis, $\epsilon_0$ the vacuum permittivity, and $R$ the intermolecular distance. When approaching each other, molecules in the $|+\rangle$ state mutually align the orientation of their dipoles and repel each other \cite{karman2018microwave}. This is illustrated by the dressed intermolecular potentials shown in Fig. 1e. The repulsion prevents the molecules from reaching short range and suppresses loss from inelastic collisions \cite{karman2018microwave}.  The shielding efficiency is limited by residual loss channels, such as tunneling of the molecule pair through the microwave barrier, reaching short range, as well as non-adiabatic transitions to other scattering channels between $|+\rangle$ and $|-\rangle$ and the spectator states $|J, \ m_J\rangle = |1, \ 0\rangle$ and $|1, \ -1\rangle$, collectively labeled $|0\rangle$.

We demonstrate collisional stability for gases with $3.0(5) \times 10^4$ NaCs molecules, prepared in an optical dipole trap (ODT), with an initial peak density of $1.0(2)\times 10^{12}$ cm$^{-3}$ and an initial temperature of 750(50) nK (see Methods for details on the sample preparation). For optimal parameters of microwave shielding, the lifetime of the molecular ensembles increases from 16(2) ms to \mbox{1.0(1) s}. Fig. 2 displays lifetime data of unshielded and shielded molecules, illustrating this dramatic change. We track both the molecule number and temperature as a function of hold time in the ODT, $t_\mathrm{hold}$, as shown in Fig. 2a and b, respectively, and fit the data with a kinetic model that includes one-body, two-body and evaporative losses (see Methods).

\section{Inelastic Collisions}

\begin{figure}
    \centering
    \includegraphics[width = \columnwidth]{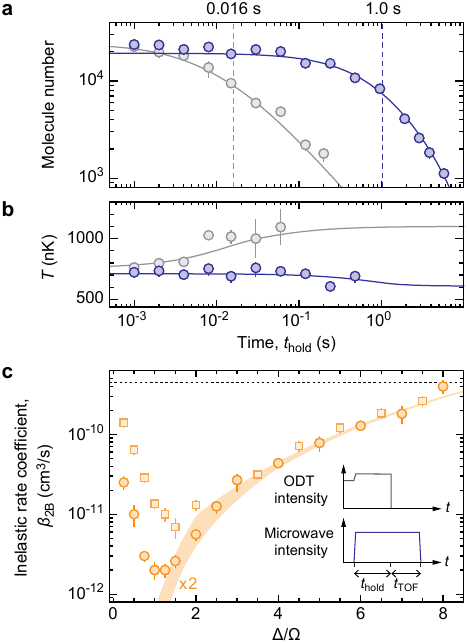}\\
    \caption{Lifetime and inelastic collisions of microwave-shielded NaCs molecules. \textbf{a}, Lifetime of molecular ensembles with (blue) and without (grey) shielding. The dashed lines indicate the respective $1/e$ lifetimes. Error bars show 1$\sigma$ standard-error-of-the-mean from ten repetitions of the measurement. The shielded data is taken at $\Omega / (2\pi) = 4$~MHz and $\Delta/\Omega = 1 $. \textbf{b}, Temperature evolution of shielded (blue) and unshielded (grey) samples, corresponding to the data in panel \textbf{a}. Error bars show the 1$\sigma$ error from the fit of the time-of-flight expansion. The solid curves in \textbf{a} and \textbf{b} are fits of the solutions of a kinetic model for molecule number and temperature. \textbf{c}, Measured inelastic rate coefficient as a function of $\Delta / \Omega$ at $\Omega / (2\pi) = 4$ MHz. Circles represent data points taken at $750 (50)$~nK and squares are taken at $160 (10)$~nK. The black dotted line corresponds to the measured two-body loss rate coefficient in the absence of microwave shielding at $750(50)$~nK. The orange shaded area shows a coupled-channel calculation for microwave ellipticities between $1^\circ$ and $5^\circ$ at 750~nK. The calculation is scaled by a factor of 2 to highlight the matching trend between experiment and theory. Insets show the relevant experimental sequence. When shielding is ramped on, the ODT power is adjusted to compensate for the change in a.c.~polarizability of the dressed state, ensuring that trap frequencies remain constant. Microwave shielding is kept on during time-of-flight to prevent inelastic losses in the initial phase of time-of-flight. Error bars show the 1$\sigma$ error from the fit of the loss curves.\\} 
    \label{fig:2}
\end{figure}

We record lifetime data under different microwave parameters, which allows us to extract the two-body loss rate coefficient, $\beta_\mathrm{2B}$, as a function of $\Delta/\Omega$. Our study is conducted at $\Omega / (2 \pi) = 4.0(4)$ MHz. This value was chosen after we measured a plateau in the shielding quality as a function of Rabi frequency between $\Omega / (2 \pi) = 4$ MHz and $\Omega / (2 \pi) = 10$ MHz, with loss rates increasing on either side of this range (see Supplementary Information Fig.~S4). The measured loss rate coefficients are shown in Fig. 2c. For $\Delta / \Omega > 1$, the trend of the data agrees well with the results of a coupled-channel calculation that takes into account our measured microwave ellipticity of $\xi = 3(2)^\circ$ (see section on Microwave Ellipticity in Supplementary Information). At $\Delta / \Omega = 1$, the data shows a 225-fold reduction in the loss rate coefficient compared to the unshielded case, going from $\beta_\mathrm{2B} = 450(50) \times 10^{-12}$ cm$^3$/s to $\beta_\mathrm{2B} = 2.0(5) \times 10^{-12}$ cm$^3$/s. For $\Delta / \Omega < 1$, the data deviates from the theoretical expectations. We compare the data taken at 750(50)~nK to a second run at 160(10)~nK. While for $\Delta / \Omega > 2$ the two runs are practically indistinguishable, there is a marked difference for $\Delta / \Omega < 2$ with a significant uptick of the inelastic rate coefficient for colder temperatures. Based on the coupled-channels calculation, which only takes into account two-body physics, such a temperature dependence is not expected. Three-body effects, not accounted for in the calculation, may be a driver of this physics, potentially in conjunction with heating caused by microwave-induced loss. A detailed understanding of these effects in future work may allow reaching even lower loss rates, as predicted by theory. 

\section{Elastic Collisions}

\begin{figure}
    \centering
    \includegraphics[width = \columnwidth]{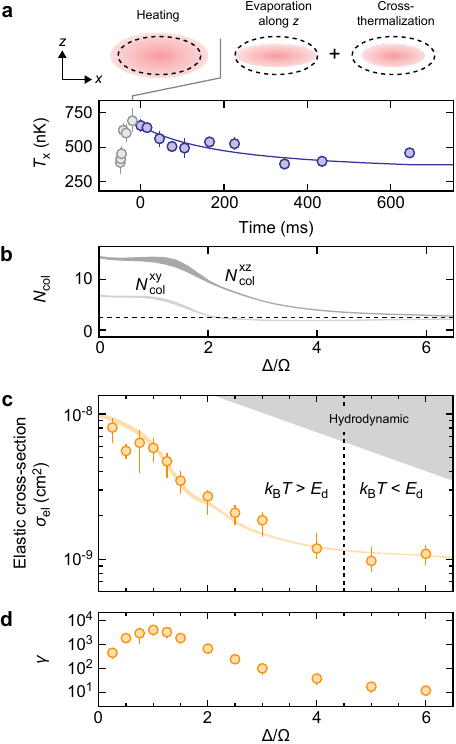}\\
    \caption{Elastic collisions of microwave-shielded NaCs molecules. \textbf{a}, Cross-thermalization experiment at $\Omega / (2\pi) = 4$ MHz and $\Delta / (2\pi) = 6$ MHz. First, shielding is turned off and two-body loss in the unshielded gas is used to heat the sample to between 650 and 800 nK via anti-evaporation \cite{ni2010dipolar} and to reduce its density; then shielding is turned back on, triggering evaporative cooling along the $z$-axis to a temperature set by the trap depth and cross thermalization with the $xy$-plane. The trap depth is kept constant during the entire sequence and the minimum trap depth is along the $-z$-direction due to gravity. Grey data points show the temperature evolution of the unshielded molecules, blue data points show the temperature in the $xy$-plane after shielding is turned back on. The blue solid line shows a fit to the data using the kinetic model. Error bars show the 1$\sigma$ error from the fit of the time-of-flight expansion. \textbf{b}, Theoretical calculation of $N_\mathrm{col}$ as a function of $\Delta/\Omega$. The dark grey (light grey) band is the calculated number of collisions for $xz$ ($xy$)-thermalization for a temperature range between 650 nK and 800 nK. The dashed line indicates $N_\mathrm{col} = 2.5$ as expected for an isotropically interacting system. \textbf{c}, Measured elastic cross section as a function of $\Delta / \Omega$, for $\Omega / (2\pi) = 4$ MHz. The orange shaded band is the integral scattering cross section from the coupled-channel calculation for a temperature range between 650 nK and 800 nK (see Methods). The grey shaded area corresponds to the hydrodynamic limit \cite{schindewolf2022evaporation}, which we do not enter at the densities used for this measurement. The dashed vertical line marks the transition between the semiclassical ($kT_\mathrm{B} > E_\mathrm{d}$) and threshold ($kT_\mathrm{B} < E_\mathrm{d}$) regimes of elastic scattering. Error bars show the 1$\sigma$ error from the fit of the thermalization curves. \textbf{d}, Ratio of elastic-to-inelastic collisions, $\gamma$, as a function of shielding parameter $\Delta/\Omega$. Error bars show the result of the error propagation in the calculation of $\gamma$.\\} 
    \label{fig:3}
\end{figure}

In addition to the suppression of inelastic collisions, the dipole moment induced by the microwave field enhances elastic collisions. The effective dipole moment depends on the microwave parameters as $d_\mathrm{eff} = d_0 / \sqrt{12 ( 1 + (\Delta / \Omega)^2 ) }$ (see Fig. 1d), and the resulting dipole-dipole interactions vary as a function of the shielding parameter, $\Delta/\Omega$. For example, for small $\Delta/\Omega$ the dressed state approaches an equal superposition of $|0,0\rangle$ and $|1,1\rangle$, leading to a maximal induced dipole moment of $d_0/\sqrt{12} \approx 1.3$ D and an enhancement of the dipolar contribution to the elastic collisions. 

As the shielding parameter $\Delta / \Omega$ is varied, the molecular gas probes two different regimes of dipolar scattering depending on the relative magnitude of the thermal energy, $k_\mathrm{B} T$, and the dipolar energy, $E_\mathrm{d} = d_\mathrm{eff}^2 / ( 4\pi\epsilon_0 \,a_\mathrm{d}^3 )$ \cite{bohn2009quasi}, where $a_\mathrm{d} = M d_\mathrm{eff}^2 / (8 \pi \hbar^2 \epsilon_0)$ is the dipolar length and $M$ the molecular mass. The elastic scattering cross section, $\sigma_\mathrm{el}$, varies in magnitude, temperature dependence, and anisotropy depending on which of these energies is dominant. For $k_\mathrm{B} T \gg E_\mathrm{d}$, collisions are semiclassical and $\sigma_\mathrm{sc} =  8\pi \, a_\mathrm{d}/ (3 k) $, where $k =  \sqrt{ \pi \, M \, k_{\mathrm{B}} T} / \hbar$ is the thermally-averaged collisional $k$-vector. For $k_\mathrm{B} T \leq E_\mathrm{d}$, i.e.~as the thermal deBroglie wavelength, $\lambda_\mathrm{th} = h / \sqrt{2\pi M k_\mathrm{B} T }$, approaches or exceeds the length scale of dipole-dipole interactions, the collisional properties are modified and the cross section enters the threshold regime, becoming $\sigma_\mathrm{th} = 32 \pi \, a_\mathrm{d}^2/ 45 + 8 \pi a_s^2$, where $a_s$ is the $s$-wave scattering length of the molecules \cite{bohn2009quasi}. Since $d_\mathrm{eff}$ is a function of $\Delta / \Omega$, our experiment accesses both regimes, with $k_\mathrm{B} T \approx E_\mathrm{d}$ at large microwave detunings and $k_\mathrm{B} T \gg E_\mathrm{d}$ close to resonance.

We measure the elastic collision cross section via a cross-thermalization experiment (Fig. 3). For these measurements we keep the peak density below $0.2 \times 10^{12}$~cm$^{-3}$ to avoid entering the hydrodynamic regime, where the thermalization rate would be limited to the mean trap frequency \cite{schindewolf2022evaporation}. At constant ODT depth, first a temperature quench is induced by turning off microwave shielding and fast heating due to two-body loss (see Methods). Then, microwave shielding is turned back on, followed by evaporative cooling and thermalization along the vertical $z$-axis and cross-thermalization with the $xy$-plane. During the sequence, we follow the temperature evolution of the cloud in $x$-direction, as shown in  Fig. 3a. Via the kinetic model (see Methods), we extract the thermalization rate $\Gamma_\mathrm{th} = \sigma_\mathrm{el}\, \bar{n} \,v_\mathrm{th} / N_\mathrm{col}$ \cite{schindewolf2022evaporation}, where $v_\mathrm{th} = 4\sqrt{k_\mathrm{B} T / (\pi M) }$ is the molecules' mean thermal velocity, $\bar{n}$ the mean density of the cloud, and $N_\mathrm{col}$ the average number of collisions required for cross-thermalization between the $z$-axis and the $xy$-plane. Smaller $N_\mathrm{col}$ means more efficient energy transfer. From $\Gamma_\mathrm{th}$ we obtain the elastic scattering cross section $\sigma_\mathrm{el}$.

The anisotropic nature of dipolar interactions has a profound impact on the thermalization dynamics via the value of $N_\mathrm{col}$. Close to resonance, where the molecular gas is in the semiclassical regime and $E_\mathrm{d}$ is small, forward collisions that do not deflect the molecules' trajectories by a large angle are favored, thus limiting the transverse energy transfer. More efficient energy redistribution is achieved in the threshold regime at larger detunings, when the induced dipole moment is lower and $E_\mathrm{d}$ becomes larger. This is shown in Fig. 3b, in which a calculation of $N_\mathrm{col}$ for cross-thermalization between the $x$ and $z$ (or $x$ and $y$) axes is provided. Unlike experiments on fermionic dipoles~\cite{aikawa2014anisotropic,li2021tuning} where there is only the dipole-dipole interaction, bosonic systems also have an $s$-wave van der Waals contribution to elastic scattering. The scattering length, $a_s$, is not known for NaCs and we obtain a value of $a_s = 1200$~$ a_0$ from a fit to our data (see Methods). With $a_s$ being the only free fitting parameter, we find excellent agreement for $\sigma_\mathrm{el}$ between the experiment and a coupled channel calculation, as shown in Fig. 3c.

From the measured elastic and inelastic collision rates, we calculate the ratio of elastic-to-inelastic collisions, $\gamma$, as shown in Fig. 3d. We observe a peak value of $\gamma \approx 4 (1) \times 10^{3}$ at $\Delta / \Omega = 1 $. The quantity $\gamma$ is typically used as a key parameter to characterize the efficacy of forced evaporative cooling \cite{davis1995analytical}. However, evaporative cooling with dipolar elastic collisions in the semiclassical regime is qualitatively different from evaporation in systems with $s$-wave or threshold dipolar interactions, as is typically the case in atomic and molecular systems, including the recent demonstrations of evaporative cooling in fermionic dipolar molecules \cite{valtolina2020dipolar,schindewolf2022evaporation}. In our case the reduced quantity $\gamma / N_\mathrm{col}$, rather than $\gamma$, sets the thermalization rate, and thus the evaporation efficiency. Our highest value of $\gamma / N_\mathrm{col} \approx 250$ is still favorable for efficient evaporation \cite{anderegg2021observation,schindewolf2022evaporation}.

\section{Evaporative Cooling}

We demonstrate evaporative cooling in the stabilized ultracold gas of NaCs molecules. We start with a gas at a temperature of $750(50)$~nK and a PSD of $5(1) \times 10^{-3}$. Then, the depth of the ODT is continuously reduced over 1.5 s, while the molecular cloud is shielded at $\Omega / (2\pi) = 4$~MHz and $\Delta / (2\pi) = 6$~MHz. At different stages of the evaporation, we measure the molecule number and temperature, and extract the phase-space density of the cloud, as shown in Fig. 4. At the end of the cooling sequence, we reach a temperature of $36(5)$~nK and a corresponding PSD of $0.10(3)$. For small molecule numbers, the measured phase-space density seems to show a plateau, likely the result of the limited signal-to-noise of our detection system. The extracted evaporation efficiency, $- d {\rm ln}({\rm PSD}) / d {\rm ln}(N)$, is $1.0(1)$. This efficiency is similar to those found in recent work on evaporative cooling of fermionic ground state molecules \cite{valtolina2020dipolar,li2021tuning,schindewolf2022evaporation}. We note that, besides the prospect of cooling the molecular gas to degeneracy, evaporative cooling allows the preparation of molecular samples at well-defined temperatures over a wide dynamic range, which will facilitate studies of quantum chemistry and collisional physics in bosonic molecular gases \cite{bause2023ultracold}. 

\begin{figure}
    \centering
    \includegraphics[width = \columnwidth]{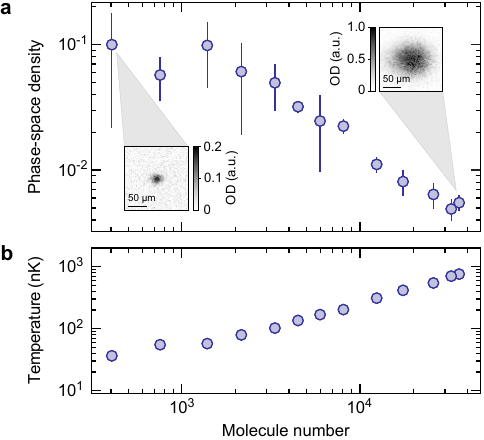}\\
    \caption{Evaporation of microwave-dressed molecules at $\Omega / (2 \pi) = 4$~MHz and $\Delta / (2\pi) = 6$~MHz. \textbf{a}, Evolution of phase-space density during evaporation. The insets show images of molecular gases after 3 ms of time-of-flight. The low temperature cloud is an average of five images; the high temperature cloud is a single image. Error bars show 1$\sigma$ standard-error-of-the-mean from three repetitions of the measurement. \textbf{b}, Temperature evolution corresponding to panel \textbf{a}. Error bars show 1$\sigma$ standard-error-of-the-mean from three repetitions of the measurement.\\} 
    \label{fig:1}
\end{figure}

\section{Outlook}

In this work, we have demonstrated that ultracold gases of bosonic ground state molecules can be effectively stabilized via microwave shielding, reaching low inelastic loss rates similar to shielded fermionic molecules despite the absence of a $p$-wave barrier. Our data on anisotropic cross-thermalization shows that, even above quantum degeneracy, the strongly dipolar character of our gas leads to nontrivial thermodynamic behavior. Dipolar liquids similar to our system have recently been predicted to show anisotropic thermal conductivity \cite{wang2022thermal} and viscosity \cite{wang2022thermoviscous}. Thanks to the rapid tunability of microwave parameters $\Omega$ and $\Delta$, allowing the quasi-instantaneous tuning of inelastic and elastic scattering properties, novel non-equilibrium measurement protocols can be envisioned to probe such thermodynamics. 

A key question that emerges from this work is whether Bose-Einstein condensation of microwave-shielded dipolar molecules can be achieved. The measured gain in phase-space density brings us to the brink of Bose-Einstein condensation. Following the observed trend in evaporation, a BEC with 100 molecules would be expected. This is currently below the signal-to-noise level of our imaging system, but straightforward upgrades will provide the necessary improvements. In addition, further reduction of microwave noise and a better understanding of inelastic loss at small detunings may allow us to reach even lower loss levels, as predicted by theory. Field-linked resonances \cite{avdeenkov2006suppression,chen2023field}, which should be accessible for NaCs at low microwave ellipticity and moderate microwave intensity \cite{chen2023field}, allow for independent tuning of $a_\mathrm{d}$ and $a_\mathrm{s}$ \cite{lassabliere2018controlling} and  offer a tuning knob to further improve on the scattering properties of our molecules. Using field-linked resonances to tailor interactions may also tune $N_\mathrm{col}$  close to or below the $s$-wave value of 2.5~\cite{davis1995analytical}, increasing the thermalization rate by an order of magnitude, enhancing our evaporation. 

Starting from shielded three-dimensional bulk samples of bosonic molecules, efficient transfer with minimal loss to lower dimensional systems comes within reach. In two-dimensional systems, microwave shielding and d.c.~electric fields should enable the realization of strongly correlated phases \cite{buchler2007strongly}, such as supersolidity and self-organized crystallization of dipoles, both in single layers \cite{schmidt2022self} and multilayers \cite{wang2006quantum}. NaCs molecules are a highly promising platform for such explorations due their large dipole moment, which allows a characteristic range of dipolar interactions, $a_\mathrm{d}$, of tens of micrometers. Shielded loading of optical lattices, in particular from a BEC of molecules, may offer a way to reach unity filling, necessary to realize extended Hubbard models \cite{Capogrosso2010,dutta2015non}, and enabling studies of many-body spin models \cite{micheli2006toolbox,gorshkov2011tunable, yan2013observation} in defect-free molecular arrays. \\

{\bf Acknowledgements.} We thank Andreas Schindewolf, Goulven Qu{\'e}m{\'e}ner, and Timon Hilker for discussions, Michal Lipson and Javad Shabani for the loan of equipment, and Tarik Yefsah for critical reading of the manuscript. This work was supported by an NSF CAREER Award (Award No.~1848466), an ONR DURIP Award (Award No.~N00014-21-1-2721), and a Lenfest Junior Faculty Development Grant from Columbia University. C.W. acknowledges support from the Natural Sciences and Engineering Research Council of Canada (NSERC). W.Y.\ acknowledges support from the Croucher Foundation. I.S. was supported by the Ernest Kempton Adams Fund. S.W.\ acknowledges additional support from the Alfred P. Sloan Foundation.\\

\setcounter{figure}{0}
\makeatletter 
\renewcommand{\thefigure}{Extended Data \@arabic\c@figure}
\makeatother

\section{Methods}

{\bf Sample preparation and detection.} NaCs Feshbach molecules are assembled from overlapping ultracold gases of Na and Cs via a magnetic field ramp across a Fesh\-bach resonance at $B_\mathrm{res} = 864.1(1)$ G \cite{lam2022high}. The magnetic field points in vertical $z$-direction and sets the quantization axis. The samples are held in a crossed optical dipole trap (ODT) with trap frequencies $\omega / (2 \pi) = (60, \ 65, \ 140)$~Hz (measured for NaCs ground state molecules). The $x$-dipole trap is elliptical and focused to waists of 127(5)~$\mu$m (horizontal) and 56(3)~$\mu$m (vertical); the $y$-dipole trap is circular with a waist of 106(5)~$\mu$m. Optical trapping light is generated by a 1064 nm narrow-line single-mode Nd:YAG laser (Coherent Mephisto MOPA).

NaCs Feshbach molecules are transferred to the electronic, vibrational, and rotational ground state, $X^1\Sigma^+ |v = 0, \ J = 0\rangle$, via stimulated Raman adiabatic passage (STIRAP) \cite{stevenson2023ultracold, warner2023pathway}. The specific hyperfine state of the molecules is $|m_{I_{\rm Na}}, \ m_{I_{\rm Cs}}\rangle = |3/2,\ 5/2\rangle$, where $m_{I_{\rm Na}}$ $(m_{I_{\rm Cs}})$ is the projection of the nuclear spin of sodium (cesium) onto the quantization axis. The STIRAP beams are pointing upward on the vertical axis. This allows NaCs molecules to undergo time-of-flight expansion while in the ground state in the presence of microwave shielding, which enables precise thermometry (see below). 

After time-of-flight expansion, NaCs molecules are detected by reversing STIRAP, optically dissociating them with a pulse of light that is resonant with the Cs $6^2 S_{1/2} |F = 3, \ m_F = 3\rangle \rightarrow 6^2 P_{3/2} |F = 4, \ m_F = 4\rangle$ transition at high magnetic field, immediately followed by absorption imaging of Cs atoms on the $6^2 S_{1/2} |F = 4, \ m_F = 4\rangle \rightarrow 6^2 P_{3/2} |F = 5, \ m_F = 5\rangle$ transition at high field.

{\bf Thermometry via time-of-flight expansion.} The temperature of NaCs ground state molecular gases is precisely measured using time-of-flight expansion. The basic methodology is similar to temperature measurements of atomic gases, where fitting the increase of cloud radius due to ballistic expansion allows the extraction of in-trap temperature \cite{ketterle1999making}. For molecules there are subtleties that need to be taken into account to faithfully measure temperature via time-of-flight expansion. The molecules cannot be directly imaged and need to be dissociated prior to imaging. Dissociation at the beginning of time-of-flight would lead to enhanced losses and systematic shifts in temperature measurement due to a momentum kick from reverse STIRAP and due to the nonadiabaticity of the reverse magnetic field ramp, as characterized in earlier work \cite{lam2022high}. Instead, we let NaCs ground state molecules expand under microwave shielding and dissociate them, right before imaging the constituent Cs atoms. The ability to perform shielding during time-of-flight prevents inelastic loss. For time-of-flight expansion of unshielded ground state molecules, we observe a systematic overestimation of temperature by about $10\%$ (see Supplementary Information Fig.~S5) which we attribute to inelastic loss in the initial phase of time-of-flight. Absorption images yield the column-density of the molecular cloud, integrated along the $z$-direction. To extract the cloud size in the $x$-direction, we integrate along the $y$-direction, fit the profile to a 1D Gaussian, $n(x) = A e^{-x^2 / (2 \sigma_x^2)}$, and extract the $\sigma_{x}$-radius at each $t_{\rm TOF}$. The temperature, $T$, is obtained from the cloud sizes by using the relation $\sigma_{x} (t_{\rm TOF}) = \sqrt{ \sigma_0^2 + (k_\mathrm{B} T/ M) t_{\rm TOF}^2}$, where $\sigma_0$ is the initial cloud radius, $k_\mathrm{B}$ is the Boltzmann constant, and $M$ is the mass of the molecule. An analogous procedure is followed for the $y$-direction, yielding a cross-check of temperature. An example for time-of-flight expansion data is shown in Extended Data Fig. 1. 

\begin{figure}
    \centering
    \includegraphics[width = 8.6 cm]{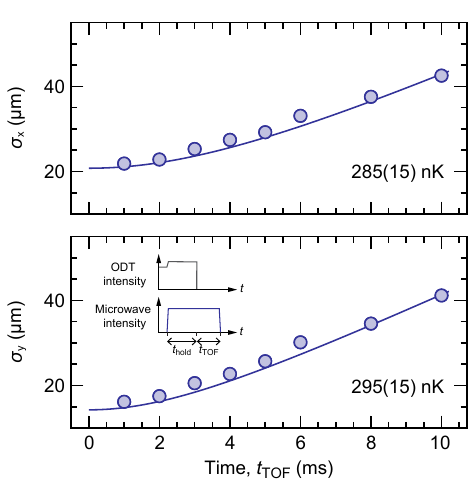}\\
    \caption{Time-of-flight expansion of a gas of ground state molecules with shielding at $\Omega/(2\pi) = 4$ MHz, $\Delta/(2\pi) = 6$ MHz. Upper (lower) panels correspond to the $x$ ($y$) direction. Insets in the lower panels illustrate the experimental sequence. The fitted mean temperature is 290(10) nK. Error bars show the 1$\sigma$ standard-error-of-the-mean from two repetitions of the measurement.}
    \label{fig:SI1}
\end{figure} 

{\bf Microwave antenna.} Circularly polarized microwave fields are generated with a phased-array microwave antenna. The antenna consists of four individual loops, arranged in a cloverleaf-shape, that are one-wavelength resonant for a frequency of 3.5~GHz. Each loop is fed by a 15~W radio-frequency amplifier (MiniCircuits ZHL-15W-422). The frequency is generated with an ultralow noise signal generator (Rohde~\&~Schwarz SMA100B) that is split into four channels via a power splitter.  Each channel is given a differential phase shift of 90$^\circ$ to generate $\sigma^+$ microwave polarization.  Before the power splitter, a voltage controlled attenuator (General Microwave D1954) allows control of microwave power and a stack of three pin-diode switches provides a 135~dB suppression of the source when off. The ellipticity of the resulting microwave is measured to be $\xi = 3(2)^{\circ}$. For further details see the section on Microwave Ellipticity in the Supplementary Information. Additional details of the microwave system are described in Ref.~\cite{yuan2023planar}.

{\bf Kinetic model.} In order to extract elastic and inelastic loss rates from lifetime data of the shielded molecular gases, we employ a fitting model that includes one-body, two-body, and evaporative losses. The following coupled differential equations describe the rate of change of molecule number and energy in the molecular gas  \cite{warner2021overlapping}:
\begin{eqnarray}
\dot{N} & = & \dot{N}_\mathrm{1B} + \dot{N}_\mathrm{2B} + \dot{N}_\mathrm{ev} \nonumber \\
\dot{E} & = & \dot{E}_\mathrm{1B} + \dot{E}_\mathrm{2B} + \dot{E}_\mathrm{ev} \nonumber
\end{eqnarray}
In our formalism, the total energy of the gas is $E = 3 N k_B T$.  The one-body terms take the usual form $\dot{N}_\mathrm{1B} = - N / \tau_\mathrm{1B}$ and $\dot{E}_\mathrm{1B} = - E / \tau_\mathrm{1B}$, where $\tau_\mathrm{1B}$ is the one-body lifetime. $\tau_\mathrm{1B}$ is measured directly by observing low density loss curves in which other losses are negligible. The measured value is as low as $\tau_\mathrm{1B} \sim 4.4(4)$ s for $\Omega = 4$ MHz (see Supplementary Information Fig.~S3) and is kept fixed at this value in the fitting model. 

The two-body term in the number differential equation is given by \cite{olson2013optimizing} $ \dot{N}_\mathrm{2B}  = -\beta_\mathrm{2B} \bar{n} N $. Here, $\beta_\mathrm{2B}$ is the two-body loss rate coefficient and $\bar{n}$ the average density of the molecular cloud. The average density is related to the peak density, $n_0$, by $\bar{n} = n_0 / (2 \sqrt{2})$, where $n_0 = N \left(\bar{\omega}^2M/(2\pi k_B T)\right)^{3/2}$, $\bar{\omega} = (\omega_x \omega_y \omega_z)^{1/3}$ the mean trap frequency, and $M$ the molecular mass. The two-body loss contribution to the energy differential equation is given by $ \dot{E}_\mathrm{2B}  = - (3/4)\beta_\mathrm{2B} \bar{n} E$. The $(3/4)$ prefactor originates from integrating the product of the energy density and the number density over the volume of the cloud. A consequence of this is that the molecular gas heats up as a result of two-body loss at a rate $\dot{T} = (1/4) \beta_\mathrm{2B} \bar{n} E/N$, giving rise to anti-evaporation \cite{ni2010dipolar}. This can be intuitively understood by noting that two-body loss preferentially takes place in the trap center where the local density is highest, while the energy per molecule in the trap center is smaller ($3k_\mathrm{B}T/2$) than the average energy per molecule in the sample ($3k_\mathrm{B}T$). Heating from anti-evaporation is used to realize the temperature quench in the measurement of the elastic scattering cross section (see Fig.~3).

The effects of evaporation are included via the term $\dot{N}_\mathrm{ev} = -N \nu(\eta) \Gamma_{\rm el} / N_{\rm col} $  \cite{luiten1996kinetic, olson2013optimizing}. Here, $\nu(\eta)$ is the fraction of elastically scattered molecules with kinetic energy higher than the trap depth and $\Gamma_{\rm el} / N_{\rm col}$ is the thermalization rate.  $\Gamma_{\rm el}$ is the elastic scattering rate, $N_{\rm col}$ is the number of collisions to produce a $1/e$ change in the molecule temperature, $\eta = U_\mathrm{min} / ( k_B T) $ is the truncation parameter and $U_\mathrm{min}$ is the trap depth. From Ref.~\cite{davis1995analytical}, $\nu(\eta) = (2 + 2 \eta + \eta^2) / (2 e^{\eta})$.  The elastic scattering rate is $\Gamma_{\rm el} = \bar{n} \sigma_\mathrm{el} v_{\rm th}$, where $\sigma_\mathrm{el}$ is the elastic scattering cross-section, and $v_{\rm th} = 4\sqrt{k_B T / (\pi M)}$ is the thermal velocity. In our fitting routine, we cap $\Gamma_{\rm el} < \bar{\omega} / ( 2 \pi )$, in order to account for the hydrodynamic limit.  As interactions in our gas are highly anisotropic, $N_{\rm col}$ is not a number, but rather a matrix accounting for the number of collisions for thermalization for every pair of trap axes, i.e. $N_{\rm col}^{\rm xx}$, $N_{\rm col}^{\rm xy}$, $N_{\rm col}^{\rm xz}$, etc.  In our fitting routine, we fit the product $\sigma_\mathrm{el} / N_{\rm col}$ and then use the calculated maximum $N_{\rm col}$ element to extract $\sigma_\mathrm{el}$.

The evaporative term in the energy differential equation is $\dot{E}_\mathrm{ev} = - (1/3) E \alpha(\eta) \Gamma_{\rm el} / N_{\rm col}$, where $\alpha(\eta) = (6 + 6\eta + 3\eta^2 + \eta^3) / (2 e^{\eta})$~\cite{davis1995analytical}.  $(1/3) E \alpha(\eta)$ is the energy of the molecules with kinetic energy larger than the trap depth and $\Gamma_{\rm el} / N_{\rm col}$ is the rate at which the energy will leave the system.

We can recast the differential equations in the form:
\begin{eqnarray}
\dot{N} & = &  -N [ 1 / \tau_\mathrm{1B} + \beta_{\rm 2B} \bar{n} + \nu(\eta) \Gamma_{\rm ev} \bar{n} ] \nonumber\\
\dot{E} & = & -E [ 1 / \tau_\mathrm{1B} + (3/4) \beta_{\rm 2B} \bar{n} + (1/3) \alpha(\eta) \Gamma_{\rm ev} \bar{n} ], \nonumber
\label{eq:de}
\end{eqnarray}
where we used $\Gamma_{\rm ev} = \sigma_\mathrm{el} v_{\rm th} / N_\mathrm{col}$ for clarity. 

Experimentally, we measure the number and temperature of the molecular cloud as a function of hold time. Then the data needs to be fitted with this model to extract the initial number, $N_0$, the initial temperature, $T_0$, $\beta_{\rm 2B}$ and $\sigma_\mathrm{el} / N_{\rm col}$.  In practice, we first fit cross-thermalization data, as shown in Fig.~3 in the main text, to obtain $\sigma_\mathrm{el} / N_{\rm col}$; then, we fit lifetime data, as shown in Fig.~2 in the main text, using $\sigma_\mathrm{el} / N_{\rm col}$ from the first fit.

{\bf Coupled-channel calculations.} We perform coupled-channel scattering calculations as described in Refs.~\cite{karman2018microwave,karman2019microwave}. Here, we give a brief description of the numerical details.

Our coupled-channels calculations describe the collision between two molecules that interact with a magnetic field, an elliptical microwave field~\cite{karman2019microwave}, and with each other through the dipole-dipole interaction.
The NaCs molecules are described as rigid rotors with $J=0,1$.
The end-over-end rotation of the molecules about one another is described by a partial wave expansion up to $L=12$.
Coupling to non-initial hyperfine states is neglected, after exploratory calculations that found no effect of hyperfine structure.
The solutions to the coupled-channels equations are propagated numerically for internuclear distances between 50 and $10^5~a_0$.
At short range, an absorbing boundary condition is imposed that models loss that occurs when the molecules come close together~\cite{karman2018microwave}.
The calculations are repeated for 31 logarithmically-spaced collision energies between 1~nK and 10~$\mu$K.
From the energy dependent cross sections we determine thermally averaged elastic and inelastic scattering rates.

In addition to overall elastic and inelastic collision rates, averaged over all incoming directions in a thermal gas, we also compute the differential cross section for elastic scattering
\begin{eqnarray}
\frac{d\sigma}{d\Omega}(\bm{k},\bm{k}') = \frac{4\pi^2}{k^2} \left| \sum_{L,M_L,L',M_L'} i^{L-L'} Y_{L',M'}(\bm{k}') T_{L',M_L';L,M_L} Y_{L,M}^\ast(\bm{k}) \right|^2, \nonumber
\end{eqnarray}
where $\bm{k}$ and $\bm{k}'$ are the initial and final wavenumber,
$Y_{L,M}(\bm{k})$ is a spherical harmonic depending on the polar angles of $\bm{k}$,
and the $T$-matrix elements are obtained from our coupled-channels calculations.
In the next section, these elastic cross sections are used to model the rate of thermalization in our dipolar gas.

{\bf Dipolar thermalization.} We follow earlier work \cite{wang2021anisotropic,gueryodelin1999collective} on thermalization in harmonically confined ultracold gases and derive equations of motion by computing moments of the Boltzmann equation
\begin{eqnarray}
\frac{d \langle q_j^2\rangle}{dt} - \frac{2}{M} \langle q_j p_j\rangle &=& 0, \nonumber \\
\frac{d \langle q_j p_j\rangle}{dt} - \frac{1}{M} \langle p_j^2\rangle + M\omega_j^2 \langle q_j^2\rangle  &=& 0, \nonumber \\
\frac{d \langle p_j^2\rangle}{dt} + 2M\omega_j^2 \langle q_jp_j\rangle &=& \mathcal{C}[\Delta p_j^2], 
\label{eq:EOM}
\end{eqnarray}
that is, equations of motion for the nine dynamical properties $\{ \langle x^2 \rangle, \langle x p_x \rangle, \langle p_x^2 \rangle, \langle y^2 \rangle,$ $ \langle y p_y \rangle, \langle p_y^2 \rangle, \langle z^2 \rangle, \langle z p_z \rangle, \langle p_z^2 \rangle \}$.
Collisions are described by the term
\begin{eqnarray}
\mathcal{C}[\Delta p_i^2] &=& \mathcal{C}_{ix} \langle p_x^2\rangle + \mathcal{C}_{iy} \langle p_y^2\rangle + \mathcal{C}_{iz} \langle p_z^2\rangle, \nonumber \\
\mathcal{C}_{ij} &=& -\frac{\bar{n}}{(Mk_BT)^2} \int d\bm{k}\ k\ c^\mathrm{eq}(k) \int d^2\Omega\ \frac{d\sigma}{d\Omega}\ \Delta\bm{k}_{i}^2\ \Delta\bm{k}_{j}^2, 
\label{eq:colintfinalij}
\end{eqnarray}
where $\bm{k}$ is the relative momentum,
$\Delta\bm{k}_{i}$ is the $i$ Cartesian component of the change in momentum,
and
\begin{eqnarray}
c^\mathrm{eq}(k) &= \frac{1}{\left( \pi M k_B T \right)^{3/2}} \exp\left(-\frac{k^2}{M k_B T}\right) \nonumber 
\end{eqnarray}
is the thermal distribution of relative momenta.

Thermalization has been studied previously for $s$-wave collisions \cite{monroe1993measurement,wu1996direct,gueryodelin1999collective},
which results in an energy-independent isotropic cross section,
and for threshold dipolar interactions \cite{wang2021anisotropic}, which results in an energy-independent but anisotropic cross section that is known analytically \cite{bohn14differential}.
The threshold dipolar results describe dipolar gases with $k_\mathrm{B} T \ll E_\mathrm{d}$.
This applies to ultracold magnetic atoms, but not necessarily to a strongly dipolar molecular gas.
For resonant dressing of NaCs the dipolar length can be as large as 40\,000~$a_0$, and the dipolar energy scale as low as 0.7~nK.
To describe thermalization in a strongly dipolar molecular gas, we here evaluate Eq.~\ref{eq:colintfinalij} numerically using elastic differential cross sections from our coupled-channels calculations.

To determine the rate of thermalization we define pseudo-temperatures $T_i = [\langle p_i^2\rangle + M^2 \omega_i^2 \langle x^2\rangle] / 2 M k_\mathrm{B}$, for each cartesian direction,
and an equilibrium temperature $T_\mathrm{eq} = (T_x+T_y+T_z) / 3$.
Then, at short times we have
\begin{eqnarray}
\frac{\partial \langle p_i^2\rangle}{\partial t} &=& \mathcal{C}_{ix} \langle p_x^2\rangle + \mathcal{C}_{iy} \langle p_y^2\rangle + \mathcal{C}_{iz} \langle p_z^2\rangle. \nonumber
\end{eqnarray}
If we bring the pseudo-temperature in the $j$ direction out of equilibrium, the pseudo-temperature in $i$ direction responds as
\begin{eqnarray}
\frac{\partial T_i}{\partial t} &=& \frac{3}{2} \mathcal{C}_{ij} \left[ T_j - T_i\right], \nonumber
\end{eqnarray}
where we used $\mathcal{C}_{ix}+\mathcal{C}_{iy}+\mathcal{C}_{iz}=0$.
Thus, at short times, the pseudo-temperatures approach equilibrium exponentially with time constant $k_{ij} = \frac{3}{2} \mathcal{C}_{ij}$.
If the collision rates, $\mathcal{C}_{ij}$, become comparable to the trap frequencies, the short-time approximation breaks down, and we instead determine the 1/$e$ thermalization time by a full simulation of the equations of motion Eqs.~(\ref{eq:EOM}) as described in Ref.~\cite{wang2021anisotropic}.

Since an overall scaling of the elastic cross section will increase the rate of both thermalization and elastic collisions,
the effectiveness of thermalization is often characterized by their ratio
\begin{eqnarray}
N_\mathrm{col}^{ij} = \frac{\bar{n} \langle v_\mathrm{th} \sigma_\mathrm{el} \rangle}{k_{ij}},
\end{eqnarray}
known as the number of elastic collisions per thermalization.
For $s$-wave collisions, this ratio is $N_\mathrm{col} = 5/2$,
whereas it has been shown that threshold dipolar collisions can lead to a smaller value of $N_\mathrm{col}$,
and indeed it has been observed for fermionic microwave-shielded NaK that $N_\mathrm{col}$ is between 1 and 2 depending on the microwave polarization \cite{schindewolf2022evaporation,chen2023field}.
We note that these measurements for NaK were in fact in the threshold regime,
and the semiclassical behavior is masked by the hydrodynamic thermalization \cite{schindewolf2022evaporation,chen2023field}.

For strongly dipolar bosonic NaCs molecules, we find substantially different behavior of $N_\mathrm{col}$, as seen in Fig.~3c of the main text.
The effect of dipolar collisions is large and can increase the number of collisions for thermalization by almost an order of magnitude above the bare $s$-wave result of $N_\mathrm{col} = 5/2$.
The increase of $N_\mathrm{col}$ results from two effects.
First, in the semi-classical regime the dipolar elastic cross section decreases as $1/\sqrt{E}$,
which emphasizes low-energy collisions that lead to less momentum transfer.
Second, in the semi-classical regime the cross section also becomes more forward scattered,
which further reduces the amount of momentum transferred.
The dipolar thermalization is also strongly anisotropic, leading to substantially different thermalization rates for collisions in and perpendicular to the plane of the microwave polarization, $N_\mathrm{col}^{xy}$ and $N_\mathrm{col}^{xz}$.\\

{\bf S-wave scattering length.} For parameters in which our bosonic molecules are in the threshold scattering regime, $k_\mathrm{B} T \leq E_\mathrm{d}$, the scattering cross-section is given by the sum of its dipolar and $s$-wave contributions, $32 \pi \, a_\mathrm{d}^2/ 45 + 8 \pi a_s^2$ \cite{bohn2009quasi}. While the dipolar part is well-defined by the microwave parameters, the $s$-wave scattering length of the NaCs molecules is not known a priori. Neither the scattering length from our coupled-channels calculations nor the universal $s$-wave scattering length, $a_s = (1-i)\bar{a}$ where $\bar{a}= [2 \pi / \Gamma(1/4)^2] (M C_6 / \hbar^2)^{1/4} \approx 580(30)$~$a_0$~\cite{julienne2011universal}, reproduce the measured elastic scattering cross section at large detunings (corresponding to $a_\mathrm{d} \approx 0$), as shown in Extended Data Fig. 2a.  
We directly add an $s$-wave contribution to the scattering matrix, which we use to fit the observed behavior at large detuning.  Our fit best agrees with the experiment for $a_s = 1200$~$a_0$.  The $s$-wave contribution affects both the scattering cross-section (dominantly at large detunings), and the anisotropy of the collisions, as shown in Extended Data Fig. 2b. \\

\begin{figure}
    \centering
    \includegraphics[width = \columnwidth]{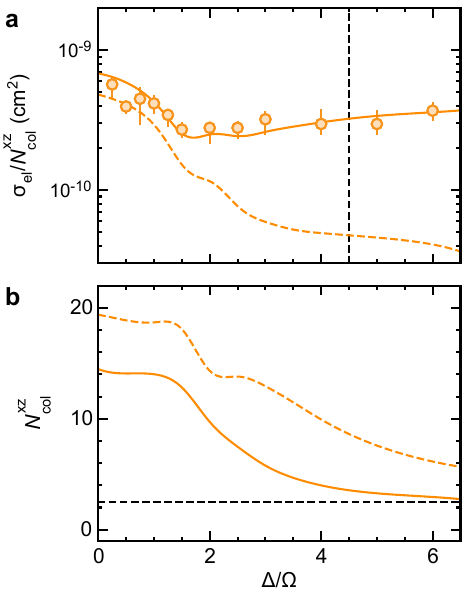}\\
    \caption{Fitting an $s$-wave scattering length.  \textbf{a}, Comparison of experimentally extracted collision cross-section, assuming $N_{\rm col} = 1$, to the calculated ratio of the elastic cross-section and $N_{\rm col}^{\rm xz}$.  Solid line corresponds to $a_s = 1200$~$a_0$ while the dashed line is obtained directly from our coupled-channels calculations. The error bars provided are the 1$\sigma$ error from the fit of the thermalization curves. \textbf{b}, Comparison of calculated $N_{\rm col}^{\rm xz}$ for $a_s = 1200$~$a_0$, orange solid line, and $a_s = \bar{a}$, orange dashed line.  The black dashed line marks $N_{\rm col} = 2.5$.}
    \label{fig:SI6}
\end{figure}

\end{document}


\title{Supplementary Information for \\ "Collisionally Stable Gas of Bosonic Dipolar Ground State Molecules"}

\author{Niccol\`{o} Bigagli}
\affiliation{Department of Physics, Columbia University, New York, New York 10027, USA}
\author{Claire Warner}
\affiliation{Department of Physics, Columbia University, New York, New York 10027, USA}
\author{Weijun Yuan}
\affiliation{Department of Physics, Columbia University, New York, New York 10027, USA}
\author{Siwei Zhang}
\affiliation{Department of Physics, Columbia University, New York, New York 10027, USA}
\author{Ian Stevenson}
\affiliation{Department of Physics, Columbia University, New York, New York 10027, USA}
\author{Tijs Karman}
\affiliation{Institute for Molecules and Materials, Radboud University, 6525 AJ Nijmegen, Netherlands}
\author{Sebastian Will}\email{Corresponding author. Email: sebastian.will@columbia.edu}
\affiliation{Department of Physics, Columbia University, New York, New York 10027, USA}

\date{\today}

\setcounter{figure}{0}
\makeatletter 
\renewcommand{\thefigure}{S\@arabic\c@figure}
\makeatother

\maketitle

\section{Quantum numbers}

We denote the internal molecular quantum states relevant to this work with labels $|J,m_J\rangle$, where $J$ is the total angular momentum of the molecules excluding nuclear spin, and $m_J$ its projection onto the quantization axis. Due to the high magnetic field in the vicinity of the Feshbach resonance at $B_\mathrm{res}= 864.1(1)$ G, the molecules are in the Paschen-Back regime and nuclear spin is decoupled from all other angular momenta. Therefore, we omit nuclear spin quantum numbers. The nuclear spin stays unchanged at $\ket{m_{I_{\rm Na}}, \ m_{I_{\rm Cs}}} = \ket{3/2,\ 5/2}$, even in the presence of the microwave field.

\section{Rotational spectroscopy}

\begin{figure} 
    \centering
    \includegraphics[width = 6.9 cm]{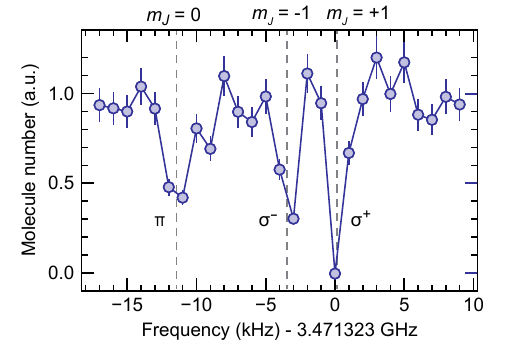}\\
    \caption{Spectrum of the $|J=0\rangle \rightarrow |J=1\rangle$ rotational transitions of NaCs using a microwave field with mixed polarization. Different polarization couples different $m_J$ states, as marked. The vertical dashed lines indicate the assigned transitions. Error bars show the 1$\sigma$ standard-error-of-the-mean from three repetitions of the measurement.}
    \label{fig:SI2}
\end{figure} 

In order to identify the $\sigma^+$ transition for microwave shielding, we performed rotational spectroscopy of NaCs. We prepared a gas of ground state molecules in the $\ket{J, \ m_J} = \ket{0, \ 0}$ state and exposed it to a microwave field with mixed polarization while the magnetic bias field was close to $B_\mathrm{res}$. The corresponding spectrum is shown in Fig.~\ref{fig:SI2}. To assign quantum numbers to each transition, we repeated the measurement with polarized microwave fields, showing a single transition depending on the polarization used. The resonance frequency of the $\sigma^+$ transition between $\ket{0, \ 0}$ and $\ket{1, \ +1}$ was measured to be 3.471323(2) GHz.  

\section{Dressed state preparation}

\begin{figure} 
    \centering
    \includegraphics[width = 6.4 cm]{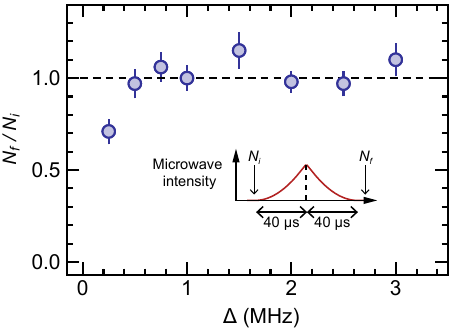}\\
    \caption{Adiabaticity of dressed state preparation for $\Omega / (2\pi) = 4$ MHz. The ratios of molecule numbers before ($N_i$) and after ($N_f$) a round trip through the dressed state $|+\rangle$ is shown. The inset shows a schematic of the experimental sequence. Error bars show the 1$\sigma$ standard-error-of-the-mean from three repetitions of the measurement.}
    \label{fig:SI3}
\end{figure} 

The molecules are prepared in the dressed state $|+\rangle$ via an adiabatic increase of the intensity of the blue-detuned microwave field. The intensity is ramped up within $40$ $\mu$s using a ramp following a quadratic power law, $\Omega(t) \propto t^2$. To confirm adiabaticity of this ramp, we compared the molecule number in state $\ket{J, \ m_J, \ m_{I_{\rm Na}}, \ m_{I_{\rm Cs}}} = \ket{0, \ 0, \ 3/2, \ 5/2}$ before the ramp, $N_i$, to the molecule number $N_f$ after a microwave ramp into and out of state $|+\rangle$. For a non-adiabatic ramp, we would expect loss of population into other states and $N_f/N_i$ should be smaller than 1. We performed this measurement for microwave Rabi frequency $\Omega/(2 \pi)=4$ MHz and various detunings $\Delta$, as shown in Fig.~\ref{fig:SI3}. While the data point at the lowest detuning does not meet this criterion, all data with $\Delta/\Omega > 0.1$, which is the case for all data in the main text, fulfills the criterion. 

\section{Microwave Ellipticity}

We measured the ellipticity of the microwave field by driving resonant Rabi oscillations on the transitions $\ket{J, \ m_J } = \ket{ 0, \ 0} $ to $\ket{ 1, \ -1 }$ and $\ket{ 1, \ +1 }$. For fixed microwave power, we determine the respective Rabi frequencies and obtain an ellipticity $\xi = \arctan(\Omega_{\sigma^-} / \Omega_{\sigma^+}) = 3(2) \degree$. An attempt to drive a resonant transition to $\ket{ 1, \ 0 }$ under identical conditions was consistent with no Rabi coupling. We note that the axis of molecular rotation is well-aligned with the axis of the magnetic field.

\section{One-body loss}

\begin{figure} 
    \centering
    \includegraphics[width = 6.4 cm]{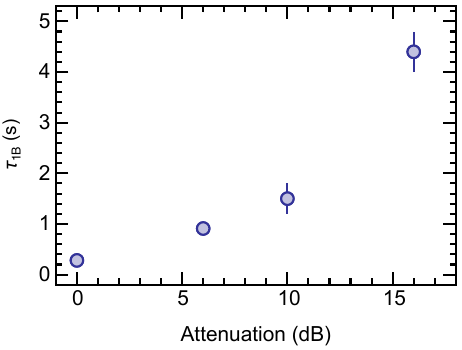}\\
    \caption{Measured one-body-limited lifetime of the shielded NaCs gas for the same Rabi frequency $\Omega / 2\pi = 4$ MHz and $\Delta/(2\pi) = 4$ MHz using different levels of attenuation of the 15 W amplifiers. The error bars show the 1$\sigma$ error from the fit of the loss curves.}
    \label{fig:SI4}
\end{figure}

\begin{figure}
    \centering
    \includegraphics[width = 6.6 cm]{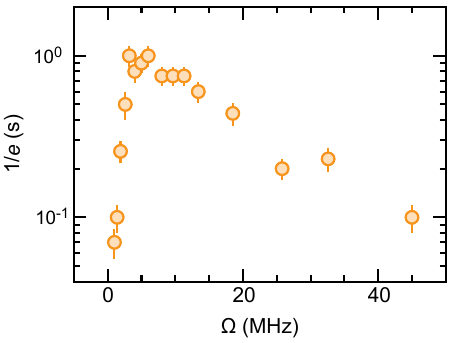}\\
    \caption{$1/e$-lifetime of microwave-shielded molecules as a function of $\Omega$. The ratio $\Delta / \Omega = 1.5$ is kept fixed. The error bars show the 1$\sigma$ error from the fit of the loss curves.}
    \label{fig:SI5}
\end{figure} 

The dominant source of one-body loss in the microwave-shielded molecular gas stems from noise of the microwave field away from the carrier frequency. Such noise can drive transitions to the anti-shielded dressed state $|-\rangle$ and unshielded spectator states $\ket{0}$, limiting the lifetime of the molecular gas. Such noise can be generated or amplified by any active component of the microwave chain, e.g.~the microwave source and amplifiers. To reduce this noise we employ a high quality microwave source (Rohde \& Schwarz SMA100B with ultralow phase-noise option). We find that the phase-noise of the source is so low that it does not limit the observed one-body loss. Instead, we find that thermal white noise of the fixed-gain microwave amplifiers (several MHz away from the carrier frequency) contributes dominantly. To reduce this noise, we set the output power of the microwave source to maximum and attenuate the amplifiers' output to the desired power level via external attenuators. We have measured one-body loss at constant overall output power, corresponding to  $\Omega / 2\pi = 4$ MHz, for different combinations of source power level and amplifier attenuation (Fig.~\ref{fig:SI4}). For the largest attenuation of 16 dB, we find the longest one-body-limited lifetime of $\tau_{1\text{B}} \sim 4.4(4)$ s.

\section{Optimum Rabi Frequency}

All the data of this work was taken for a Rabi frequency of $\Omega / 2\pi = 4$ MHz. For this Rabi frequency we were able to make use of the full 16 dB post-amplifier attenuation to limit one-body loss from microwave noise (see above). Fig.~\ref{fig:SI5} shows the measured $1/e$-lifetimes of the molecular cloud for different Rabi frequencies $\Omega$, while keeping the ratio $\Delta/\Omega = 1.5$ fixed. To access higher Rabi frequencies the post-amplifier attentuation was gradually reduced, which reduced the lifetimes due to higher microwave noise. Theoretically, higher Rabi frequency is expected to lead to better shielding than lower Rabi frequency. Within the technical limitations of the experimental setup, we observe peak shielding performance for $\Omega / 2\pi \sim 4$ MHz.

\section{Thermometry via Time-of-flight expansion}

\begin{figure*}
    \centering
    \includegraphics[width = 14.2 cm]{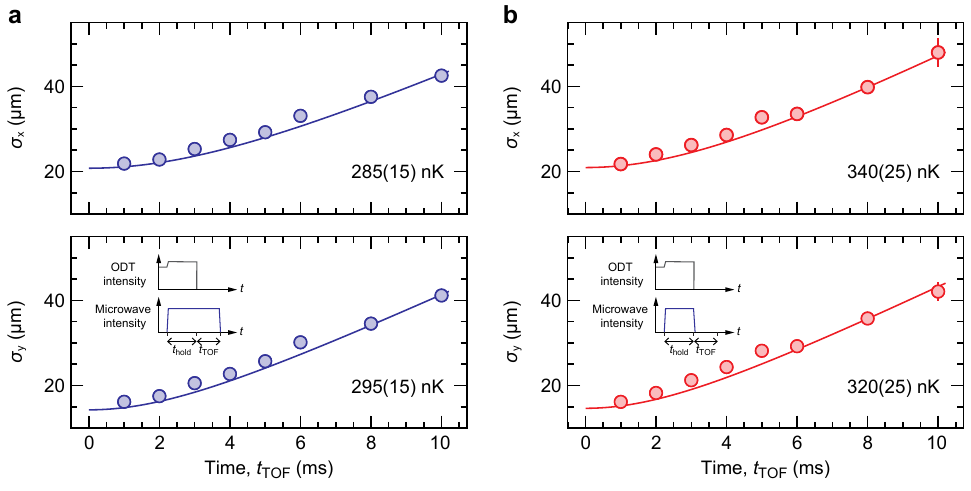}\\
    \caption{Time-of-flight expansion of a gas of ground state molecules with shielding at $\Omega/(2\pi) = 4$ MHz, $\Delta/(2\pi) = 6$ MHz (a) and without shielding (b). Upper (lower) panels correspond to the $x$ ($y$) direction. Insets in the lower panels illustrate the respective experimental sequence. The fitted mean temperature is 290(10) nK for (a) and 330(20) nK for (b). Error bars show 1$\sigma$ standard-error-of-the-mean from three repetitions of the measurement.}
    \label{fig:SI1}
\end{figure*}

Precise thermometry of the ultracold gas of ground state molecules requires careful consideration. In the Methods section we point out a systematic shift between the temperatures measured for the time-of-flight expansion for shielded and unshielded molecules. Unshielded molecules undergo rapid loss especially at the beginning of time-of-flight expansion when the density is still high (with a two-body loss rate of up to $10^{-9}$ cm$^3$/s) which leads to a systematic overestimation of temperature by $10\%$. A comparison of time-of-flight expansion for shielded and unshielded molecules is shown in Fig. \ref{fig:SI1}. All thermometry in this work is performed using time-of-flight expansion of shielded molecules.

